\documentclass[%
   amsmath,amssymb,
   aps,
   prb,
   twocolumn
  ]{revtex4-1}

\usepackage{graphicx}
\usepackage{color}
\usepackage{bm}
\usepackage{acro}
\usepackage{xspace}
\usepackage{multirow}
\usepackage[percent]{overpic}

\DeclareAcronym{BZ}{
  short=BZ,
  long=Brillouin zone,
  }
\DeclareAcronym{TWS}{
  short=TWS,
  long=topological Weyl semimetal
  }
\DeclareAcronym{WS}{
  short=WS,
  long=Weyl semimetal
  }
\DeclareAcronym{TDS}{
  short=TDS,
  long=topological Dirac semimetal
  }
\DeclareAcronym{WN}{
  short=WN,
  long=Weyl node
  }
\DeclareAcronym{3D}{
  short=3D,
  long=three-dimensional,
  }
\DeclareAcronym{2D}{
  short=2D,
  long=two-dimensional,
  }
\DeclareAcronym{SOI}{
  short=SOI,
  long=spin-orbit interaction,
  }
\DeclareAcronym{SOC}{
  short=SOC,
  long=spin-orbit coupling,
  }
\DeclareAcronym{SO}{
  short=SO,
  long=spin-orbit,
  }
\DeclareAcronym{ARPES}{
  short=ARPES,
  long=angular resolved photoemission spectroscopy,
  }
\DeclareAcronym{DFT}{
  short=DFT,
  long=density-functional theory,
  }
\DeclareAcronym{XC}{
  short=XC,
  long=exchange and correlation,
  }
\DeclareAcronym{PBE}{
  short=PBE,
  long={Perdew, Burke and Ernzerhof},
  }
\DeclareAcronym{GGA}{
  short=GGA,
  long=generalized-gradient approximation,
  }
\DeclareAcronym{PW}{
  short=PW,
  long=plane wave,
  }
\DeclareAcronym{EELS}{
  short=EELS,
  long=electron energy loss spectra,
  }
\DeclareAcronym{bct}{
  short=bct,
  long=body-centered tetragonal,
  }
\DeclareAcronym{QE}{
  short=QE,
  long=\texttt{Quantum ESPRESSO},
  }
\DeclareAcronym{DOS}{
  short=DOS,
  long=density of states,
  }
\DeclareAcronym{IR}{
  short=IR,
  long=infrared,
  }
\DeclareAcronym{UV}{
  short=UV,
  long=ultraviolet,
  }
\DeclareAcronym{TaAs}{
  short=TaAs,
  long=tantalum arsenide,
  }
\DeclareAcronym{scf}{
  short=scf,
  long=self-consistent,
  }
\DeclareAcronym{HWCC}{
  short=HWCC,
  long=hybrid Wannier charge centers,
  }
\DeclareAcronym{FS}{
  short=FS,
  long=Fermi surface,
  }
\DeclareAcronym{TR}{
  short=\emph{T},
  long=time-reversal,
  }
\DeclareAcronym{INV}{
  short=\emph{I},
  long=inversion,
  }
\DeclareAcronym{FM}{
  short=FM,
  long=ferromagnetic,
  }
\DeclareAcronym{AFM}{
  short=AFM,
  long=antiferromagnetic,
  }
\DeclareAcronym{QHE}{
  short=QHE,
  long=quantum Hall effect,
  }
\DeclareAcronym{AQHE}{
  short=AQHE,
  long=anomalous quantum Hall effect,
  }

\newcommand{\angstrom}{\textup{\AA}}

\newcommand{\kgrid}[1]{$#1 \times #1 \times #1$}

\begin{document}
  \title{The type-I antiferromagnetic Weyl semimetal InMnTi$_2$}

  \author{Davide Grassano}
  \email{davide.grassano@epfl.ch}
  \affiliation{Theory and Simulations of Materials (THEOS), and National Center for Computational Design and Discovery of Novel Materials (MARVEL), \'Ecole Polytechnique F\'ed\'erale de Lausanne, CH-1015 Lausanne, Switzerland}

  \author{Luca Binci}
  \affiliation{Theory and Simulations of Materials (THEOS), and National Center for Computational Design and Discovery of Novel Materials (MARVEL), \'Ecole Polytechnique F\'ed\'erale de Lausanne, CH-1015 Lausanne, Switzerland}

  \author{Nicola Marzari}
  \affiliation{Theory and Simulations of Materials (THEOS), and National Center for Computational Design and Discovery of Novel Materials (MARVEL), \'Ecole Polytechnique F\'ed\'erale de Lausanne, CH-1015 Lausanne, Switzerland}
  \affiliation{Laboratory for Materials Simulations (LMS), Paul Scherrer Institut (PSI), CH-5232, Villigen PSI, Switzerland}


  \begin{abstract}
    Topological materials have been a main focus of studies in the past decade due to their protected properties that can be exploited for the fabrication of new devices.
    Among them, Weyl semimetals are a class of topological semimetals with non-trivial linear band crossing close to the Fermi level.
    The existence of such crossings requires the breaking of either \ac{TR} or \ac{INV} symmetry and is responsible for the exotic physical properties. In this work we identify the full-Heusler compound InMnTi$_2$, as a promising, easy to synthesize, \acs{TR}- and \acs{INV}-breaking \acl{WS}. To correctly capture the nature of the magnetic state, we employed a novel $\mathrm{DFT}+U$ computational setup where all the Hubbard parameters are evaluated from first-principles; thus preserving a genuinely predictive \textit{ab initio} character of the theory. We demonstrate that this material exhibits several features that are comparatively more intriguing with respect to other known \aclp{WS}:
    the distance between two neighboring nodes is large enough to observe a wide range of linear dispersions in the bands,
    and only one kind of such node's pairs is present in the \acl{BZ}.
    We also show the presence of Fermi arcs stable across a wide range of chemical potentials.
    Finally, the lack of contributions from trivial points to the low-energy properties makes the materials a promising candidate for practical devices.
  \end{abstract}

  \maketitle

  \acresetall

  \section{Introduction}
    Topological semimetals\cite{armitage2018weyl} constitute a class of materials where protected band crossings occur.
    They can be distinguished as either Dirac\cite{wang2012dirac} or \acp{WS}\cite{wan2011topological,murakami2007phase}, when the crossings happens at isolated points in the \ac{BZ}, or nodal line semimetals\cite{burkov2011topological}, when the crossings span an entire line.
    For the former case, the crossings, also called nodes, can be assigned to be Dirac or Weyl with a $k \cdot p$ framework, which can take the form of either a Dirac or a Weyl Hamiltonian.
    The former describes a fourfold degenerate crossing with no chirality, while the latter describes a pair of twofold degenerate crossings with opposite chiralities, as guaranteed by the fermion doubling theorem\cite{NIELSEN1981219}.
    The presence of these nodes lead to the emergence of quasi-particle excitations with a behavior similar to that of a Dirac or Weyl fermion\cite{pal2011dirac}, respectively.
    The focus of this paper is on \acp{WS}, which exhibit a wide range of interesting properties such as the Adler-Bell-Jackiw anomaly\cite{adler1969axial,bell1969pcac} related to the observation of negative magneto-transport\cite{ghimire2015magnetotransport,zhang2016signatures,arnold2016negative,gooth2017experimental}, and the presence of atypical surface states known as Fermi arcs\cite{wan2011topological}.
    \acp{WS} have been proposed for many applications ranging from the realization of q-bits\cite{castelvecchi2017strange}, to Veselago lenses\cite{hills2017current} and lasing\cite{oktay2020lasing}.

    One of the main requisites for the realization of a \ac{WS} is the breaking of either \ac{TR} or \ac{INV} symmetry.
    If both are present at the same time, it can be shown that two nodes with opposite chirality will always be degenerate, giving rise to a four-fold crossing with zero Chern number.
    A major breakthrough in the study of \acp{WS} was achieved with the experimental realization of the \ac{INV}-breaking TaAs\cite{huang2015weyl,lv2015experimental} and the other transition-metal monopnictides\cite{Belopolski.Xu.ea:2016:PRL,sun.wu.2015}.
    Unfortunately, these materials also display a complex band-structure landscape that can obfuscate the Weyl properties.
    This can happen due to several features, from the proximity of two neighboring nodes in the same pair which causes a distortion of the expected linear dispersion, to the presence of trivial points\cite{grassano2020influence}.
    In the search for possible \acp{WS} also several \ac{TR}-breaking materials have been proposed, such as the pyrochlore iridates\cite{wan2011topological}, \ac{AFM} Mn$_3$Sn and Mn$_3$Ge\cite{nakatsuji2015large,yang2017topological}, or a promising family of cobalt-based Heusler compounds \cite{wang2016time,kubler2016weyl}.
    In \ac{TR}-breaking \acp{WS}, the restriction of having \acp{WN} appearing in multiples of 4 is lifted, and the number of pairs present can range from a single to multiple ones.
    Recently, an interesting paradigm in the design of \acp{WS} has been proposed with the introduction of the \ac{FM} ternary compounds RAlX [R = rare earth, X=Ge,Si]\cite{chang2018magnetic}, like CeAlSi\cite{yang2021noncollinear}, where both \ac{TR} and \ac{INV} symmetries are broken.
    In this case the material can be seen as an \ac{INV}-breaking \ac{WS} with a Zeeman-like perturbation that causes a shift in the position of the nodes depending on their Chern number.
    These materials have the same crystal structure as the transition-metal monopnictides, and indeed exhibit the same 4/8 pairs of W1/2 nodes, shifted depending on their Chern number and, in some cases, depending on the material composition, also additional W3 and W4 nodes, always organized in $4n$ pairs.

    In this letter we perform a systematic study of the properties of InMnTi$_2$, a non-centrosymmetric full-Heusler compound\cite{graf2009crystal} belonging to space group 216.
    Among several possible spin configurations, we predict that the \ac{AFM} phase is the most stable one.
    As alluded to above, we expect to find Weyl nodes organized in $4n$ pairs with a Zeeman-induced shift in their positions depending on the node's Chern number.
    Indeed, in the present analysis, we find 12 pairs of type-I nodes\cite{soluyanov2015type} throughout the entire \ac{BZ} of the material, which are all equivalent modulus a minor shift in $k$-space induced by the the \ac{AFM} phase.
    We show that in InMnTi$_{2}$ the distance between a pair of neighboring nodes is greater than that in \ac{INV}-breaking \acp{WS} such as the transition-metal monopnictides, and greater or comparable to that of \ac{TR}-breaking ones, while still being very close to the Fermi level.
    Finally, we also show that only a trivial pocket of electrons is present at $\Gamma$ in addition to the Fermi pockets given by the \acp{WN}, and its contribution is negligible to the Weyl properties, as shown in the computed \ac{DOS} and optical properties.

  \section{Methods}
    \Ac{DFT} calculations have been carried out using the open-source \acl{QE}\cite{Giannozzi.Baroni.ea:2009:JoPCM,espresso2017} distribution in combination with the norm-conserving full-relativistic pseudopotentials from the ONCVPSP library\cite{hamann2013optimized}, with the \acl{XC} functional derived within the \acl{PBE} \acl{GGA}\cite{PBE.1996}.
    The parameters and convergence thresholds set for the calculations are higher than conventional ones, in order to guarantee an accuracy of the order of $1$~meV for the resulting band eigenvalues.
    In particular, we use a \acl{PW} cutoff for the wavefunctions \textbf{G}-vectors of 150~Ry.
    The initial relaxation and self-consistent calculations are carried out using a \kgrid{9} Monkhorst-Pack mesh to sample the \ac{BZ}.
    Successively, a non self-consistent calculation with a \kgrid{12} grid has been used to perform the Wannierization of the wavefunctions with Wannier90\cite{pizzi2020wannier90}.
    In order to determine the magnetic configuration of the material, several supercells have been considered with different starting spin configurations.
    To account for the localized nature of $3d$ electrons in transition-metals, we employed the fully relativistic noncollinear parametrization of Dudarev's $\mathrm{DFT}+U$ functional \cite{Dudarev1998,dudarev2019}. The internal consistency and the predictive character of the method are guaranteed by directly evaluating the Hubbard parameters for the Ti$-3d$ and Mn$-3d$ states also including spin-orbit coupling. To do this, we used the linear response \cite{cococcioni2005linear} density-functional perturbation theory approach \cite{baroni2001,Timrov2018}, that we recently generalized to the noncollinear fully relativistic case \cite{binci2023}. In this scheme, the Hubbard $U$ correction is given by $U^I=(\chi^{-1}_0-\chi^{-1})_{II}$, where the interacting (noninteracting) response matrix $\chi$ ($\chi_0$) are given by the curvature of the total energy with respect to an external field constraining the $3d$ orbital occupation: $\chi_{II'}=\sum_i^\mathrm{occ}\big(\langle\Psi_i|P^I|\delta_{I'}\Psi_i\rangle+\langle\mathcal{T}\Psi_i|\mathcal{T}P^I\mathcal{T}^\dagger|\mathcal{T}\delta_{I'}\Psi_i\rangle\big)$. The linearized perturbed wavefunctions satisfy two Sternheimer equations; a standard and a time-reversed one \cite{binci2023}:
    \begin{gather}
    \Big(H_{[\mathbf{B}]}-\epsilon_i\Big)|\delta_{I}\Psi_i\rangle=-\mathcal{P}\;\delta_{I} V_\mathrm{KS}^{[\mathbf{B}]}|\Psi_i\rangle\\
    \Big(H_{[-\mathbf{B}]}-\epsilon_i\Big)|\mathcal{T}\delta_{I}\Psi_i\rangle=-\Pi\;\delta_{I} V_\mathrm{KS}^{[-\mathbf{B}]}|\mathcal{T}\Psi_i\rangle
    \end{gather}
    where $\mathbf{B}_\mathrm{xc}$ is the magnetic exchange correlation potential. With this method, using a 4$\times$4$\times$4 \textbf{q}-points mesh, we estimated $U=2.45$~eV for the Mn$-3d$ and $U=2.27$~eV for the Ti$-3d$ states.
    
    The bands used for the Wannierization have been chosen so as to include the entire isolated manifold including both the valence and conduction bands around the Fermi level. 
    Following the Wannierization, WannierTools\cite{WU2017} has been employed to perform a Wannier interpolation on a denser \kgrid{151} k-point mesh, and to find the position of all points with a gap smaller than $0.5$~meV.
    The position of the crossings has then also been verified with direct \ac{DFT} calculations.
    The denser grids from WannierTools have been used for the calculation of the Fermi surface and \ac{DOS}.
    The iterative Green's function method\cite{guinea1983effective,lee2005band} is used in order to compute the surface states on a tetragonal supercell.
    The surface is cut along the $[001]$ direction and the states are computed both for the Ti/Mn- and Ti/In-terminated surfaces
    
    The Berry curvature $\Omega$ defined as the curl of the Berry connection ($A$)
    \begin{subequations}
    \begin{align}
      \label{eq:bconn}
      A_n(\bf{k}) & = i \langle
          u_{n{\bf k}}
          |
          {\mathbf\nabla}_{\bf k}
          |
          u_{n{\bf k}}
        \rangle.
      \\
      \label{eq:bcurv}
      \mathbf{\Omega}_n (\bf{k}) & = \nabla_{\bf{k}} \times A_n(\bf{k})
    \end{align}
    \end{subequations}
    has been computed using WannierBerri\cite{tsirkin2021high}.
    The Chern number of the nodes has been computed both by considering the flux of $\mathbf{\Omega}_n (\bf{k})$ on a sphere surrounding the node \eqref{eq:chern_int}, and using \texttt{Z2pack}\cite{gresch2017z2pack} to track the evolution of the \acp{HWCC} on a sphere surrounding each node.
    \begin{align}\label{eq:chern_int}
      C = \int_\mathrm{BZ} d\mathbf{S} \cdot \bm{\Omega}_{n} (\bf{k})
    \end{align}

    The optical conductivity of the material has been computed within the independent-particle approach\cite{Adolph.Gavrilenko.ea:1996} using the Kubo-Greenwood formula as implemented in Wannier90.
    The diagonal elements of the real part of the optical conductivity are given by\cite{Bechstedt:2015:Book,Grosso.PastoriParravicini:2000:Book}
    \begin{align}\label{eq:fgr}
    \begin{split}
      \sigma _{1,jj} (\omega) =& \frac{2\pi e^2}{m^2\omega V} \sum_{\bf k} \sum_{c,v} \left[f\left(\varepsilon_v({\bf k})\right)-f\left(\varepsilon_c({\bf k})\right)\right]
      \\
      & \times
      \left|\langle c{\bf k}|p_j|v{\bf k}\rangle\right|^2\delta\left(\varepsilon_c({\bf k})-\varepsilon_v({\bf k})-\hbar\omega\right)
    \end{split}
    \end{align}
    where $V$ is the cell volume and $v$ and $c$ are the valence and conduction bands indexes respectively.
    In order to ensure that the optical properties are converged within a broadening of 3~meV, we perform the calculation on a set of increasingly denser $k$-points meshes up to a \kgrid{300} grid, where the convergence criteria is satisfied.
    The imaginary part of the dielectric tensor can be derived from the optical conductivity using the formula
    \begin{align}
      {\rm Im}\,\varepsilon_{jj} (\omega) = \frac{4\pi}{\omega}{\rm Re}\,\sigma_{jj}(\omega)
    \end{align}

    The entire process has been automated with AiiDA\cite{pizzi2016aiida,huber2020aiida} workflows, which let us keep track of the metadata and provenance for every step of the calculation.
    The relative plugins for \ac{QE} and \texttt{Z2pack} have also been employed.

  \section{Results and discussions}
    InMnTi$_2$ has been selected from a high-throughput screening performed on materials derived from the ICSD and COD databases, in which the material was previously identified as a Weyl semimetal with a non-magnetic ground state \cite{grassano2023weylscreen,mcloud2023weyl}.
    When testing magnetic configurations, even when considering supercells, the non-magnetic one is always the most energetically favored at the \ac{PBE} level.
    On the contrary when including Hubbard corrections, the non-magnetic state becomes metastable, and the lowest energy state becomes the \ac{AFM} one in the primitive cell, where the in which the Ti and Mn atoms form 2 sublattices with opposite spin configuration.
    The cell parameter derived from the relaxation is the same as that of the original screening of 3.134~\angstrom.

    \begin{figure*}
      \includegraphics[width=0.98\textwidth]{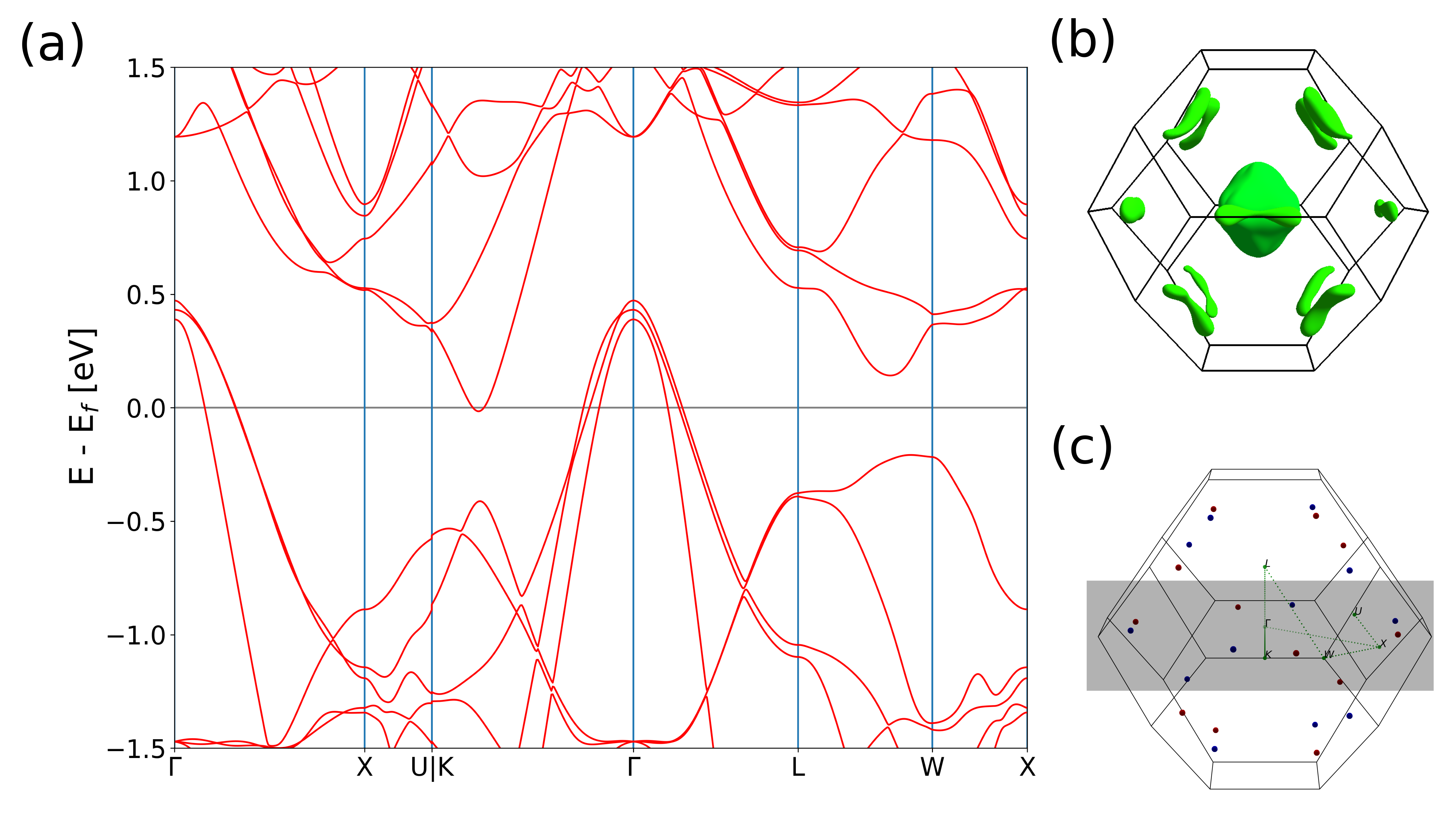}
      \caption{
        (a) Band structure along the high-symmetry path.
        \label{fig:bands}
        (b) Plot of the \ac{FS} obtained from a Wannier interpolation on a \kgrid{151} k-point grid.
        The \ac{FS} shows pockets surrounding the 12 pairs of \acp{WN}.
        Only one other pocket of trivial points is present around $\Gamma$.
        \label{fig:fermisurface}
        (c)  Plot of the \ac{BZ} including the position of the \acp{WN} in the antiferromagnetic ground state.
        The nodes with chirality +1 are shown in red, while the one with chirality -1 are shown in blue.
        The green lines show the high-symmetry path on which the bands have been calculated.
        \label{fig:bz}
      }
    \end{figure*}
    It is worth noting that the material has also been predicted to be a \ac{WS} by chemical substitution and study of the anomalous Nernst effect by Noky et al.\cite{noky2018strong}, starting from the work of Shi et al\cite{shi2018prediction}, in which they study an equivalent inverted Heusler compound Ti$_2$MnAl.
    It should be stated that for this material they obtain the antiferromagnetic configuration as the most stable one, with opposite spin directions on Ti and Mn sublattices, already at the \ac{PBE} level.

    The band dispersion of the material in the antiferromagnetic configuration is shown in Fig.~\ref{fig:bands}a, from which several features can be observed.
    Several bands are crossing at the $\Gamma$ point, which results in the presence of a trivial pocket in the Fermi surface (Fig.~\ref{fig:fermisurface}b).
    Also, a direct gap can be observed along the $\Gamma \rightarrow K$ direction, which is a direct consequence of the presence of a pair of \acp{WN} in its proximity.
    Across the entire \ac{BZ}, 12 pairs of \acp{WN} can be identified, as shown in Fig.~\ref{fig:bz}c.
    All the nodes can be mapped to a single node position $k_W = (0.2361, 0.3712, 0.0000)$ $96$~meV below the Fermi level, plus a Chern and $K$ dependent momentum shift, as expected from an \ac{INV} and \ac{TR} breaking Weyl semimetal.
    The difference between the band structure and node positions between the DFT+U \ac{AFM} results and the non-magnetic PBE ones is shown is the supplementary material (see Fig.~SM2 and SM3).
    Notably, the shift for the nodes in the $k_z = 0$ plane is of $0.0047$~\angstrom$^{-1}$ along the $k_z$ direction with sign opposite to the Chern number of the node (see Fig.~\ref{fig:bz}).
    The energy of the nodes also remain the same within the accuracy of 1~meV. 

\begin{figure*}
      \includegraphics[width=0.98\textwidth]{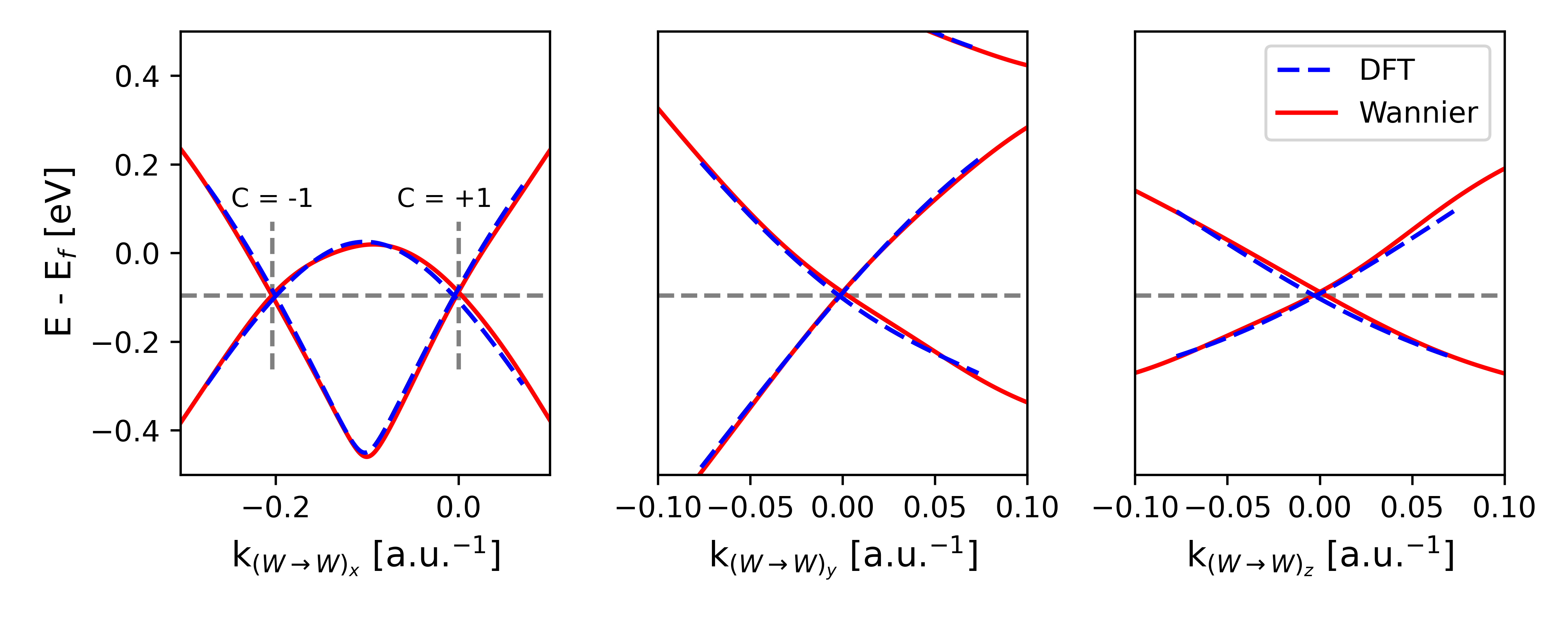}
      \caption{
        Plot of the DFT (dashed blue) and Wannier (red) band dispersion in proximity of a Weyl node.
        The $W \rightarrow W$ $x$ direction is set to line that joins a pair of adjacent Weyl nodes.
        The $y$ and $z$ directions are chosen to be perpendicular to $x$ and each other with $y$ laying in the $xy$ plane.
        \label{fig:XYZ}
      }
    \end{figure*}

    We then compute the band dispersion along the line connecting two adjacent \acp{WN} $(W \rightarrow W)_x$ and two direction $y$ and $z$ perpendicular to it and to each other, with $y$ belonging to the $xy$ plane (see Fig.~\ref{fig:XYZ}).
    This has been performed both in \ac{DFT} and with Wannier interpolations, in order to show that the Wannier function are able to reproduce the band dispersions of the nodes with a 1~meV accuracy.
    When dealing with pair of nodes, we are interested in the distance between each other as it is tied to the linear range of the band dispersion and hence the strength of the Weyl character of the nodes.
    The separation between nodes is also directly related to the possible strength of the \ac{QHE} in the Weyl semimetal\cite{yang_quantum_2011}.
    In the case of the node pairs in InMnTi$_2$ their distance is 0.37~\angstrom$^{-1}$ which is roughly four times the spacing present in the W2 nodes of TaAs and 10 time that of the W1 ones\cite{grassano2018validity}.
    This large distance for a pair means that the linear dispersion of a single node holds for a range of $60$~meV, allowing for the low-energy properties tied to the Weyl fermion picture to clearly manifest themselves in the material.
    Another two parameters that can be derived from the dispersion are the Fermi velocity tensor and the tilt vector\cite{grassano2020influence}.
    The latter can also be used to classify the \ac{WN} as of type-I or II\cite{soluyanov2015type}.
    Using this analysis for InMnTi$_2$ we observe that the nodes are of type-I, given that the tilt of the node is not strong enough.
    The fact that the nodes are of type-I can also be seen by observing the plot of the Fermi surface in Fig.~\ref{fig:bz}b, where closed electron pockets are clearly visible around each pair of nodes.
    Also a trivial hole pocket is present around $\Gamma$, as expected from the band structure.
    We show in the later discussion of the \ac{DOS} and optical properties that the presence of this trivial pocket does not give a noticeable contribution to the low-energy properties of the material, in contrast to other \acp{WS} like the transition-metal monopnictides\cite{grassano2020influence}.
    We also hypothesize that band structure engineering such as strain, electric field or doping could be used to bring the bands at $\Gamma$ below the Fermi level while not altering the topology of the material\cite{grassano2018detection}.
    \begin{figure}
      \includegraphics[width=0.98\columnwidth]{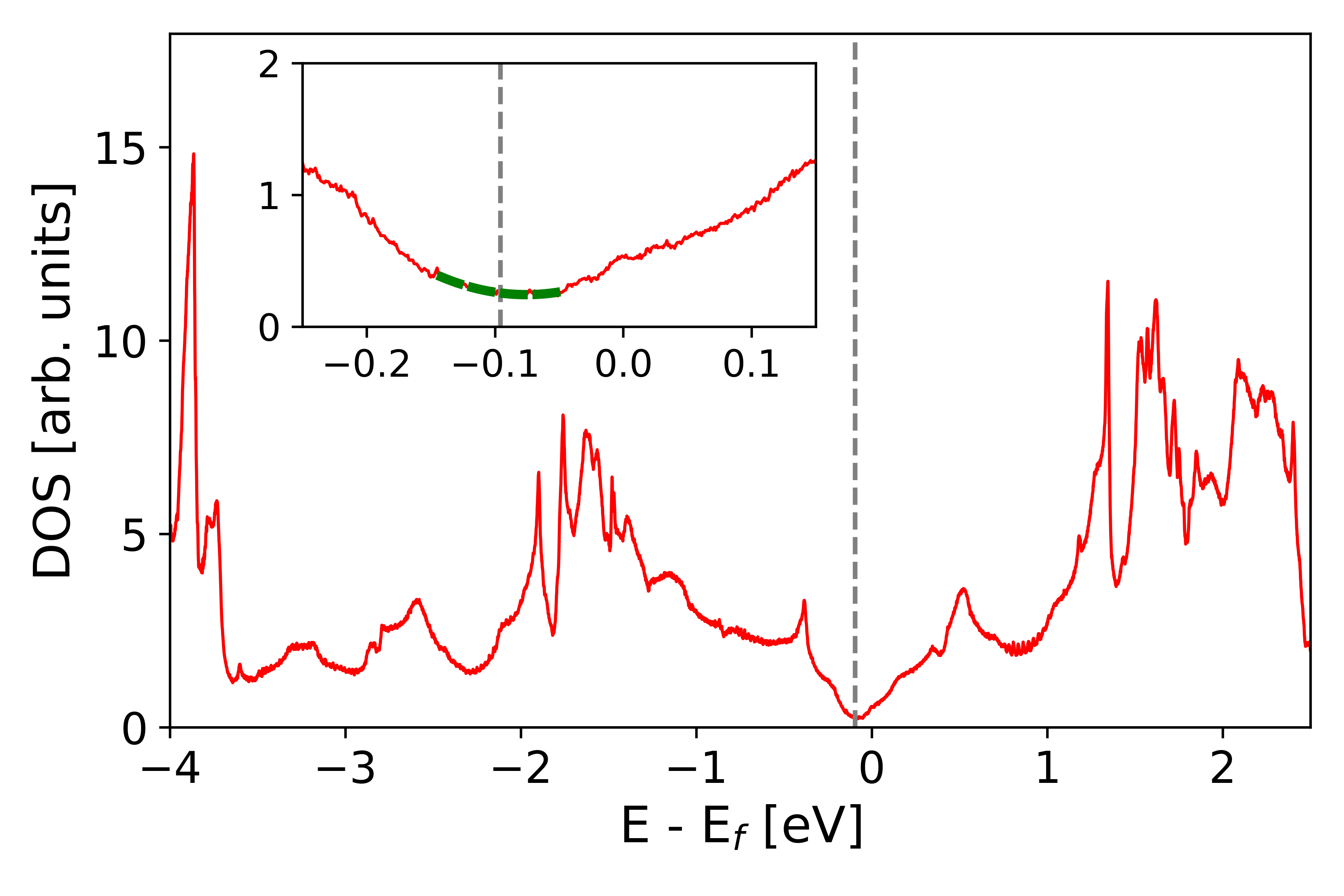}
      \caption{
        Plot of the \ac{DOS} near the Fermi level computed from a Wannier interpolation on a \kgrid{151} k-point grid with a gaussian smearing of 1~meV.
        The inset shows a zoom in an energy range closer to the Fermi level;
        in green is the fit of the \ac{DOS} near the \ac{WN} using a quadratic polynomial for the expected 60~meV range.
        The \ac{DOS} has a minimum at -0.096~meV which is the expected position of the \acp{WN} with respect to the Fermi level.
        The fact that the \ac{DOS} does not go to zero is the consequence of the presence of the pocket at $\Gamma$, but the expected quadratic behavior is still clearly present. 
        \label{fig:dos}
      }
    \end{figure}
    By analyzing the \ac{DOS} around the node position $E_{W}$ (see Fig.~\ref{fig:dos}), we can show that the presence of the trivial pocket at $\Gamma$ will have a negligible contribution to the Weyl properties.
    Indeed, given a three-dimensional \ac{WN}, the expected behavior of the \ac{DOS} should be $D(E) \sim (E - E_{W})^2$\cite{grassano2020influence} in a region around $E_{W}$.
    We show that it is possible to perform a quadratic fit around the node position (inset of Fig.~\ref{fig:dos}), in the same range as the linearity range derived from the node band dispersion, with a p-value greater than 0.99.
    The only deviation from the ideal behavior is that the \ac{DOS} does not go exactly to zero at $E_{W}$, as a consequence of the presence of the trivial pocket at $\Gamma$.
    \begin{figure}
      \includegraphics[width=0.98\columnwidth]{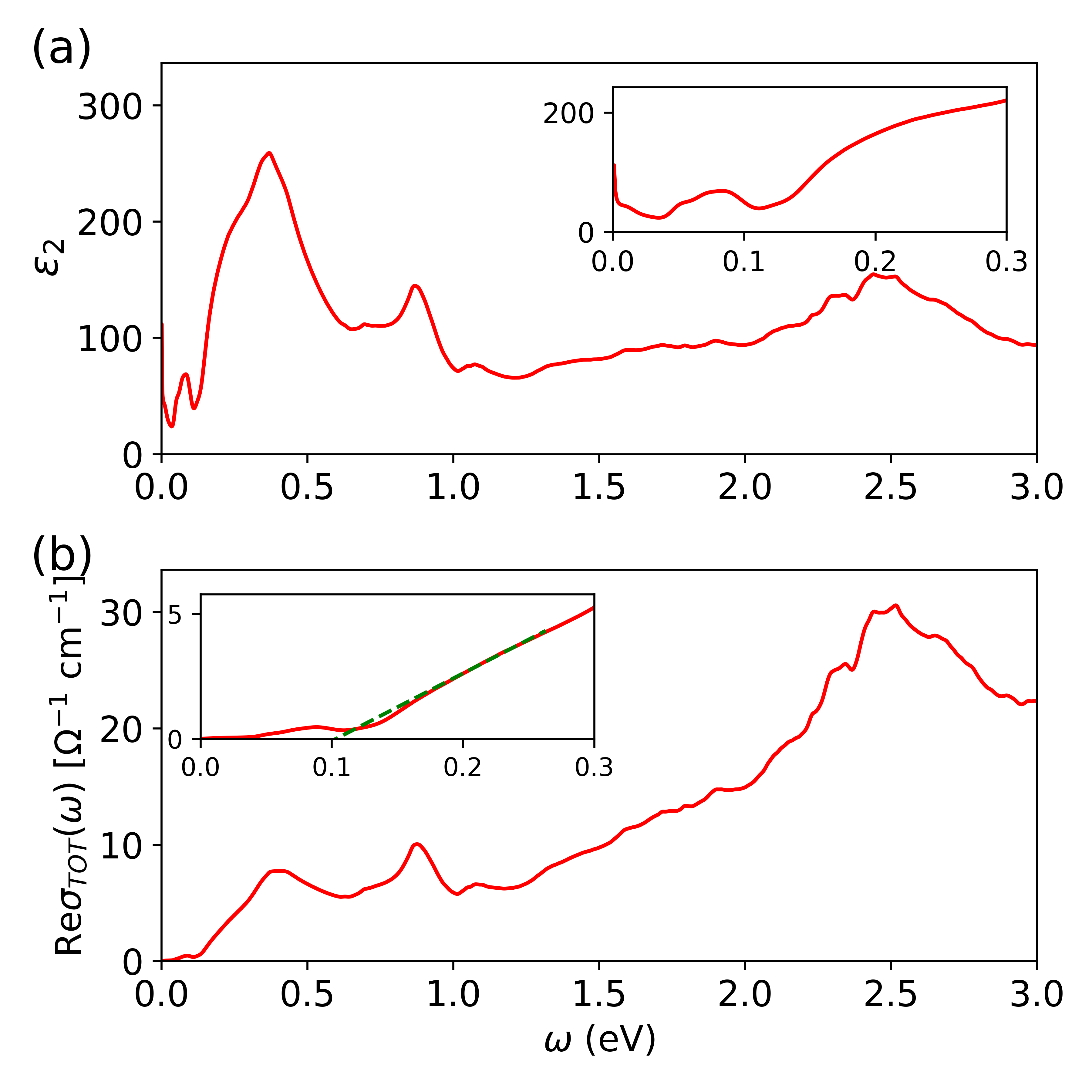}
      \caption{
        (a) Plot of the computed imaginary part of the dielectric function.
        (b) Plot of the real part of the optical conductivity.
        In green, the results of the linear fit starting from the onset energy given by $2 E_w$.\\
        The plots show that the low-energy excitation properties fully agree with the model for a type-I \ac{WS}.
        Indeed, we observe that real part of the optical conductivity can be fitted with a linear dispersion starting from an onset of 150~meV in a range wider than the expected 60~meV. 
      }\label{fig:epsTOT}
    \end{figure}
    The fact that the low-energy properties are perfectly compatible with a Weyl fermion picture can also be observed from the optical conductivity, shown in Fig.~\ref{fig:epsTOT}b.
    For a type-I \ac{WN} the real part of the optical conductivity is expected to have a linear dispersion in the low energy range ($\omega \rightarrow 0$), starting from a onset of $2 E_{W}$ due to the \acp{WN} not being exactly aligned with the Fermi level\cite{grassano2018validity}.
    In Fig.~\ref{fig:epsTOT}b we show that a linear fit can be performed starting from an onset of 150~meV.
    This value is slightly lower than the expected 192~meV due to the slight anisotropy of the \ac{WN} dispersion along the $k _{(W \rightarrow W)_y}$;
    still we observe a linear dispersion of the optical conductivity in a range wider than the expected 60~meV.

\begin{figure*}
      \includegraphics[width=0.98\textwidth,height=0.85\textheight, keepaspectratio]{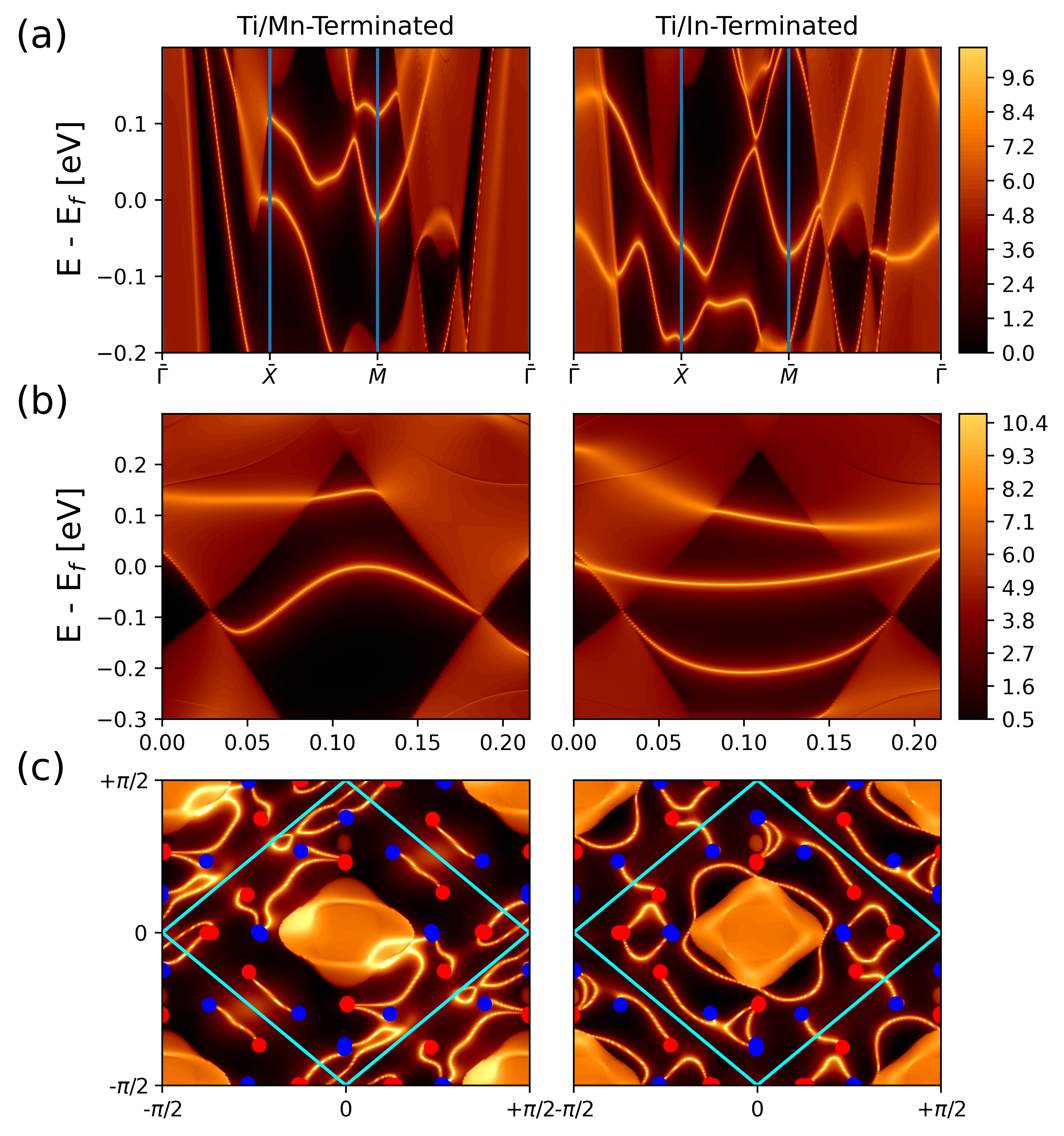}
      \caption{
        Surface states computed with the iterative Green's function method on top of the Wannierization for both the titanium and indium terminated surfaces cut along the $[001]$ direction.
        (a) \ac{ARPES} plot along the high-symmetry path in proximity of the Fermi level
        (b) \ac{ARPES} plot along the $W \rightarrow W$ direction.
        The touching point of the Weyl nodes from the bulk states is clearly visible, as well as the more pronounced bands related to the Fermi arcs that joins the 2 nodes.
        (c) Fermi surface of the surface states computed at $\mu=-0.096$.
        The red(blue) dots represent the projection of the Weyl nodes with chirality +1(-1).
        THe Fermi arcs going from nodes of opposite chirality can be observed.
        \label{fig:SurfaceStates}
      }
    \end{figure*}

    We then proceed to compute the surface states of the material, for a tetragonal supercell cut along the $[001]$ direction.
    The spectral function of both the Ti/Mn- and Ti/In-terminated surfaces along the high-symmetry lines of the surface \ac{BZ} are shown in Fig.~\ref{fig:SurfaceStates}a.
    A clear distinction between the bulk and the surface states can be inferred by the intensity of the plots.
    In Fig.~\ref{fig:SurfaceStates}b we compute the same quantity along the direction connecting the projection of two Weyl nodes with opposite chirality.
    Here we can observe the projection of the same bulk states that are also shown in Fig.~\ref{fig:XYZ}, plus a surface band leaking from one bulk node into the other which is directly related to the Fermi arcs.
    Finally, in Fig.~\ref{fig:SurfaceStates}c we show a plot of the 2D Fermi surface at the chemical potential of $\mu=-0.96$~meV, corresponding to the position of the \acp{WN}.
    As expected, we observe the presence of open lines commonly known as "Fermi arcs"  connecting the projections of the pair of nodes, highlighted in the figure with red/blue dots.
    The Fermi arcs are clearly defined, being separated from the bulk states, and can be longer than $1$~\angstrom$^{-1}$, which should make them easily detectable experimentally.
    The presence of seemingly double arcs in the Ti/In-terminated surface is due to the overlap of the projections 2 nodes at the $\pm k_z$ positions.
    The movies provided as supplementary materials also show the evolution of the arcs shape by varying the chemical potential from -100 to 0 meV.
    From this we can see that the arcs remain stable across a wide range of chemical potentials, even when the Fermi pockets around the nodes, or other trivial surface states start to interact/cross with them.

\begin{figure}
      \includegraphics[width=0.98\columnwidth]{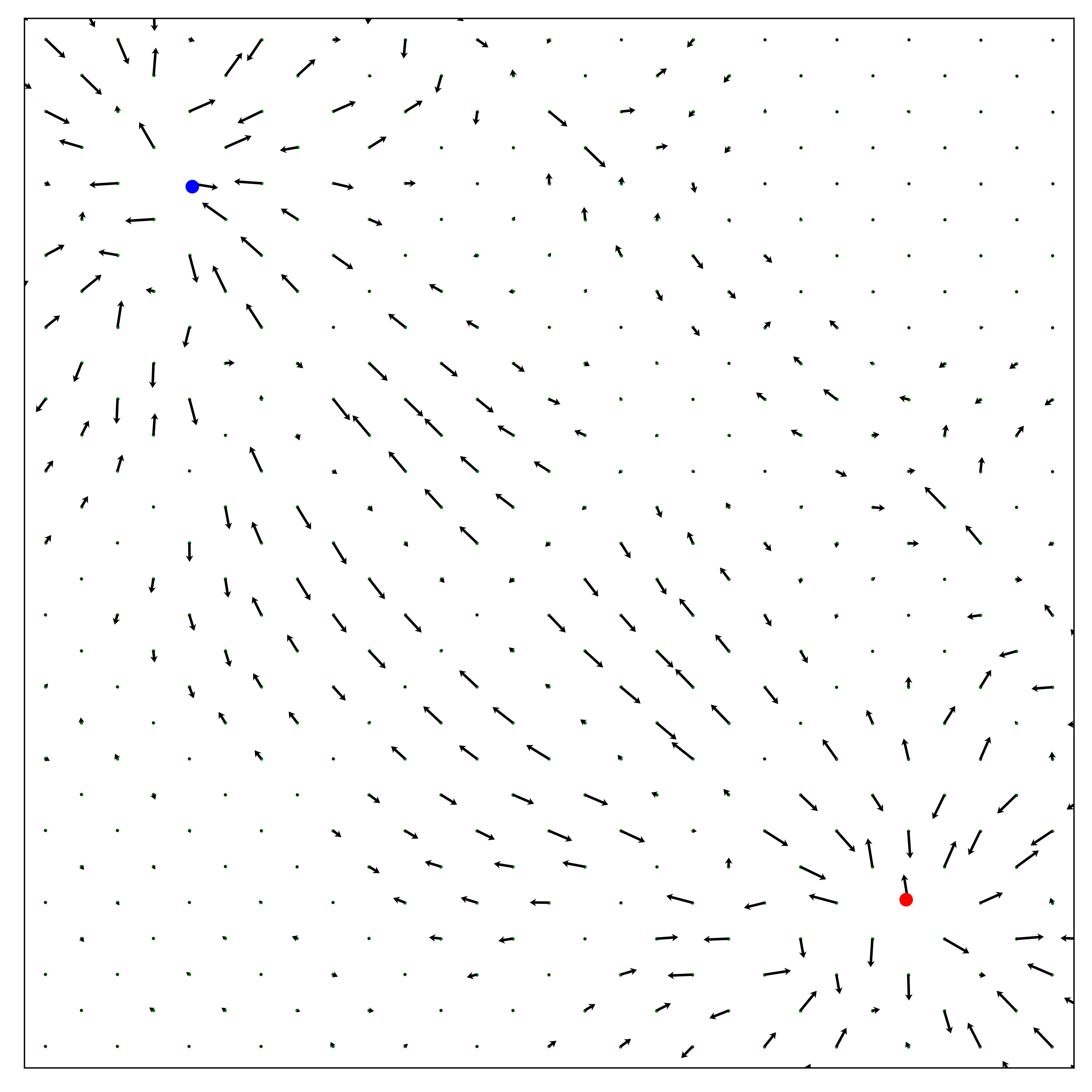}
      \caption{
        Plot of the Berry curvature on a $k_z = 0$ plane in proximity of 2 neighboring \acp{WN}.
        The red and blue dots represent a node with chirality +1 and -1 respectively.
        While the 2D plot highlights only the in-plane components of the Berry curvature, the magnitude along the $z$ axis is also taken into account in order to give a more realistic picture when trying to infer the chirality of a node from the plot.
      }\label{fig:Bcurvature}
    \end{figure}

    Finally, we compute the Berry curvature by means of Wannier interpolation on a dense regular grid of k-points.
    We can then interpolate the value of the Berry curvature on a custom grid, such as a sphere around a \ac{WN} which can than be used to derive the Berry curvature flux, the integral of which \eqref{eq:chern_int} will give the Chern number associated with the node.
    The analysis shows that two neighboring nodes belonging to a pair are of opposite chirality, as we would expect.
    The results from the study of the evolution of the \acp{HWCC} corroborates the previous results, showing that the two nodes are of chirality +1 and -1 respectively.
    We also show in Fig.~\ref{fig:Bcurvature} the Berry curvature computed on the $k_z = 0$ plane in proximity of a pair of \acp{WN};
    while the general trend shows the Berry curvature going from the $C=+1$ node to the $C=-1$ one, this is not universally true as one would expect from an ideal pair model\cite{grassano2020influence}.

  \section{Conclusions}
    In conclusion we suggest that InMnTi$_2$ is a type-I antiferromagnetic Weyl semimetal with properties that would make it an excellent candidate for future experimental studies of low-energy Weyl fermion physics, due to the presence of only one kind of Weyl point with excellent pair separation and linearity range.
    We also show that, even if a trivial pocket is present at $\Gamma$, it does not contribute significantly to the low-energy properties, and could furthermore be eliminated via material band engineering.
    These characteristic reflects on the quality of the observable surface states.
    In particular, we predict the presence of Fermi arcs that are clearly defined, with a length of over 1 \angstrom$^{-1}$, and well separated from the bulk states on a wide range of chemical potentials.

  \section{Acknowledgements}
    This research was supported by the NCCR MARVEL, a National Centre of Competence in Research, funded by the Swiss National Science Foundation (grant number 182892).

  \section{Competing Interests}
    The Authors declare no Competing Financial or Non-Financial Interests.

  \section{Data Availability}
    The data supporting the findings of this paper are available on the Materials Cloud\cite{talirz2020materials} at 
    Ref.~\onlinecite{mcloud2023inmnti2}.

  \section{Author contributions}
    D.G. is responsible for the conceptualization of the project.
    D.G. performed the DFT, Wannierization and post-processing calculations.
    L.B. provided the noncollinear $\mathrm{DFT}+U$ code and performed the DFPT calculations to derive the \textit{ab initio} Hubbard parameters.
    N.M. supervised the project.
    All authors contributed to the discussions and writing of the manuscript.

  \bibliography{bibliography}

\end{document}